\documentclass[letterpaper]{article}  
\usepackage{aaai24}  
\usepackage{times}  
\usepackage{helvet}  
\usepackage{courier}  
\usepackage[hyphens]{url}  
\usepackage{graphicx} 
\usepackage{natbib}  
\usepackage{caption} 
\usepackage{amsmath}
\usepackage{algorithm}
\usepackage{algorithmic}
\usepackage{booktabs}
\usepackage[table,xcdraw]{xcolor}
\usepackage{multirow}

\usepackage{xcolor}

\usepackage{newfloat}
\usepackage{listings}
\DeclareCaptionStyle{ruled}{labelfont=normalfont,labelsep=colon,strut=off} 
\floatstyle{ruled}
\newfloat{listing}{tb}{lst}{}
\floatname{listing}{Listing}

\usepackage{amsmath}
\usepackage{tcolorbox}
\usepackage{enumitem}
\tcbuselibrary{listings}
\usepackage{tcolorbox}
\usepackage{afterpage}
\usepackage{cellspace}
\usepackage{subfigure}

\usepackage{pbox}

\newtcblisting{promptbox}[1]{%
  title=#1,
  listing only,
  colback=white!95!gray,
  colframe=gray,
  width=\columnwidth,
  arc=0mm,
  boxrule=0.5mm,
  colbacktitle=white!95!gray,
  coltitle=black,
  fonttitle=\bfseries,
  left=6pt,
  right=6pt,
  top=6pt,
  bottom=6pt,
  listing options={basicstyle=\small\ttfamily, breaklines=true}
}

\title{Narratives of Collective Action in YouTube's Discourse on Veganism}

\author {
    Arianna Pera,\textsuperscript{\rm *}
    Luca Maria Aiello\textsuperscript{\rm $\dagger$}
}
\affiliations {
    IT University of Copenhagen\\
    $^*$arpe@itu.dk, $^\dagger$luai@itu.dk
}

\begin{document}

\maketitle

\begin{abstract}
Narratives can be powerful tools for inspiring action on pressing societal issues such as climate change. While social science theories offer frameworks for understanding the narratives that arise within collective movements, these are rarely applied to the vast data available from social media platforms, which play a significant role in shaping public opinion and mobilizing collective action. This gap in the empirical evaluation of online narratives limits our understanding of their relationship with public response. In this study, we focus on plant-based diets as a form of pro-environmental action and employ Natural Language Processing to operationalize a theoretical framework of moral narratives specific to the vegan movement. We apply this framework to narratives found in YouTube videos promoting environmental initiatives such as Veganuary, Meatless March, and No Meat May. Our analysis reveals that several narrative types, as defined by the theory, are empirically present in the data. To identify narratives with the potential to elicit positive public engagement, we used text processing to estimate the proportion of comments supporting collective action across narrative types. Video narratives advocating social fight, whether through protest or through efforts to convert others to the cause, are associated with a stronger sense of collective action in the respective comments. These narrative types also demonstrate increased semantic coherence and alignment between the message and public response, markers typically associated with successful collective action. Our work offers new insights into the complex factors that influence the emergence of collective action, thereby informing the development of effective communication strategies within social movements.
\end{abstract}

\section{Introduction}

Storytelling is a cornerstone of modern communication, with its influence permeating diverse realms from marketing to political discourse~\cite{lund2018power, seargeant2020art}. The power of compelling narratives extends beyond individual persuasion and holds the potential to catalyze collective shifts in opinion, mobilize consensus, and foster social cooperation toward shared goals. Social media have democratized opportunities for creating and sharing narratives on a global scale, empowering individuals to shape opinion formation, democratic deliberation, and collective action~\cite{yasseri2016political, jennings2021social, monti2022language}. In particular, video-sharing websites like YouTube have emerged as powerful platforms for disseminating information and mobilizing communities~\cite{uldam2013online}.

In the past decade, social media facilitated the rapid emergence of social movements that weave narratives in favor of environmental protection and climate action. Research suggests that climate discussions on social platforms, especially those related to initiatives that focus on daily life activities, can significantly enhance public awareness of environmental issues~\cite{mavrodieva2019role}. One such activity is the transition to a plant-based diet, which is seen as a tangible and effective action to mitigate the environmental damage caused by over-consumption of animal-based products~\cite{masson2019climate, judge2022dietary}. Over the past decade, pro-vegan and pro-vegetarian movements have launched various \emph{challenges} to raise awareness and encourage the public's transition towards a plant-based diet. A prominent example is \emph{Veganuary}, an initiative started in 2014 encouraging people to embrace the vegan diet for the entire month of January each year. In 2023, the initiative attracted over 700k official participants worldwide~\cite{Veganuary}. The two other largest initiatives of this kind, spanning over a month each year, are \emph{Meatless March} and \emph{No Meat May}. The success of these challenges relies heavily on participants and activists sharing their experiences online, as this generally fosters a sense of community and increases engagement~\cite{hou2023sharing}.

The effectiveness of these pro-environmental campaigns is thought to be closely linked to the types of narratives that they employ~\cite{fernandez2015analysing}. The framing of environmental issues can influence attention and shape participation in political debates~\cite{nisbet2015framing}, for instance impacting the reactions to environmental scandals~\cite{torelli2020greenwashing}. Incorporating persuasive communication strategies into pro-environmental messages can shape the public perceptions of climate change and facilitate behavioral shifts~\cite{pelletier2008persuasive}. However, despite the extensive literature on storytelling in social movements~\cite{davis2002narrative, fiskio2012apocalypse, napoli2020vegan} and the widely reported presence of collective action in climate change discourse~\cite{bamberg2015collective, van2019hope, schmitt2019predicts, hamann2020my, furlong2021social, gulliver2021assessing, judge2022dietary, suitner2023rise}, there is very little quantitative evidence of how much different narratives on climate change covered by theoretical research are effectively used in social media, and how strongly they are associated with a public response that pledges support to climate action. Addressing this gap is crucial to assess the applicability of existing theoretical frameworks to common social media practices and to enhance our understanding of the persuasive potential of different narrative types.

In an effort to bridge this knowledge gap, we delved into an exploration of vegan movements and their social campaigns as a form of pro-environmental action. We operationalized a state-of-the-art theoretical framework of narratives that classifies stories pertinent to the vegan ideology using moral foundations~\cite{napoli2020vegan}. We applied this framework to YouTube videos featuring plant-based transition challenges (such as \emph{Veganuary}, \emph{Meatless March}, and \emph{No Meat May}) to categorize the narratives they present. We did so to answer the following research questions:

\vspace{2pt} \noindent \textbf{RQ1.} \emph{How closely do narratives used in YouTube discussions about plant-based diets align with existing theoretical categories?}

\vspace{2pt} \noindent \textbf{RQ2.} \emph{Are certain narrative types more frequently associated with a public response advocating for collective action?}

\vspace{2pt} \noindent \textbf{RQ3.} \emph{What are the factors that can explain the emergence of collective action responses in the reactions to videos promoting plant-based diets?}
\vspace{2pt}

Our study's contribution lies in (i) operationalizing an existing social science theoretical framework of moral narratives used in vegan discourse and mapping YouTube video transcripts to it, (ii) examining the extent to which the language used in video comments elicits the concept of collective action, and (iii) evaluating the significance of various factors in explaining the emergence of collective action language markers in comments.

Our findings indicate that videos whose narratives promote social fight, either in the form of protesting or attempting to convert others to the cause, tended to attract reactions characterized by a more frequent presence of linguistic markers that hint at collective action. These videos, along with those emphasizing freedom of choice, demonstrated the highest levels of semantic consistency across creators, and the highest semantic alignment between the video transcript and the respective comments. Overall, we found that an increased frequency of mentions of collective action in video reactions was significantly associated with a smaller size of the crowd of commenters, a greater content coherence of a video within the reference narrative cluster, and a higher frequency of linguistic markers expressing \emph{loyalty}, a moral dimension that characterizes narratives about social fight, freedom of choice, and duty to educate others through inspirational examples.

\section{Theoretical framework}
Social science literature provides various theoretical frameworks for understanding the role of storytelling in the context of social movements (see Related Work). While such movements originate from societal changes occurring at a macro scale, their micro-level emergence is shaped by cognitive processes, often triggered by collective action \emph{frames}~\cite{johnston2005frames}. These frames serve as structures that shape individuals' perceptions and motivations within a movement. \emph{Narratives} can be seen as the embodiment of these frames, used by opinion leaders as persuasive tools. 

Given our specific interest in plant-based diets as a response to climate change, we narrow our focus to narratives within the vegan movement. Within this narrowed scope, scholars have primarily relied on qualitative studies and interviews to infer common narrative patterns. Through an in-depth analysis of vegan blogs,~\citet{napoli2020vegan} found six narrative types that are described with a combination of moral foundations and identity framing that emerge from the stories. Compared to other theoretical categorizations of vegan narratives types~\cite{waters2022v,aavik2023going}, the framework by~\citet{napoli2020vegan} is semantically more comprehensive, and offers an operationalizable map within the context of online content. Therefore, we adopt such a narrative framework as our theoretical reference, focusing on the vegan movement as a specific case of activism.
While the framework predominantly emphasizes animal welfare over explicit climate change concerns, its broad scope allows for a versatile shift towards generic environmental protection without sacrificing relevance. 
In the remainder of the paper, we will refer to the framework as the \emph{moral vegan ideology}. Next, we will describe the framework itself, outline its theoretical underpinnings, and propose a way to operationalize it.

\subsection{Veganism narratives}
\begin{table}[t!]
\centering

\begin{tabular}{c|c|c}
& \textbf{\begin{tabular}[c]{@{}c@{}}Communal\\ oriented\end{tabular}} & \textbf{\begin{tabular}[c]{@{}c@{}}Agency\\ oriented\end{tabular}} \\ \hline
\textbf{\begin{tabular}[c]{@{}c@{}}Sanctity of \\ life\end{tabular}}  
& \begin{tabular}[c]{@{}c@{}}\textit{Duty to care}\\(innocence loss,\\harm to animals,\\confronting\\content)\end{tabular}                                         & \begin{tabular}[c]{@{}c@{}}\textit{Right to good health}\\(\textbf{personal} battles,\\health \textbf{benefits},\\compassionate\\relationship)\end{tabular}      \\ \hline
\textbf{\begin{tabular}[c]{@{}c@{}}Enacting the\\ authentic\\self\end{tabular}} 
& \begin{tabular}[c]{@{}c@{}}\textit{Duty to educate}\\(\textbf{inform},\\\textbf{inspire} the\\community,\\status remark)\end{tabular}  
& \begin{tabular}[c]{@{}c@{}}\textit{Right to inner peace}\\(self-acceptance,\\tension with\\pre-vegan life)\end{tabular}   \\ \hline
\textbf{Freedom}    & \begin{tabular}[c]{@{}c@{}}\textit{Duty to fight}\\(\textbf{convert} others,\\\textbf{protest}/boycott,\\harm to animals)\end{tabular}                                                        & \begin{tabular}[c]{@{}c@{}}\textit{Right to choose}\\(\textbf{personal} story,\\\textbf{discuss} diversity\\of meanings,\\respect)\end{tabular}                                                   
\end{tabular}
\caption{Six narratives types introduced by~\citet{napoli2020vegan}, defined as a combination of identity expression (columns) and moral foundations (rows). Each narrative is divided into sub-categories shown in parenthesis. Sub-categories highlighted in \textbf{bold} are those that we found empirically in YouTube videos through our experiments.}
\label{tab:narratives_categories}
\end{table}

The \emph{moral vegan ideology} framework delineates two groups of archetypal narratives based on the type of collective identity they express: \emph{communal-oriented} and \emph{agency-oriented}. This distinction is based on whether concern for others or self-interest is featured more prominently when presenting the efforts required for driving social change. This partition aligns with the dichotomous moral positioning observed in empirical psychological studies~\cite{frimer2009reconciling}, and with established dimensions of social cognition~\cite{cuddy2008warmth}. Each of these two groups contains three narrative types that are characterized by distinct moral foundations integral to the vegan ideology: sanctity of life, enacting the authentic self, and freedom. This categorization results in the following narrative types: \emph{duty to educate}, \emph{duty to care}, and \emph{duty to fight} for the communal-oriented identity, and \emph{right to good health}, \emph{right to inner peace}, and \emph{right to choose} for the agency-oriented identity. Table~\ref{tab:narratives_categories} provides a summary of the six narratives and their associated sub-categories extracted from the original theoretical formulation of the framework.

The concept of morality intertwines with the distinct narrative types within a specific collective identity group. The \emph{duty to care} and \emph{right to good health} narratives, both linked to the sanctity of life foundation, embody respectively the collective imperative to prioritize animal welfare alongside that of humans and the value of personal health and well-being. In the pursuit of enacting the authentic self, the \emph{duty to educate} narrative strives to inform and inspire others, particularly through community networks, while the \emph{right to inner peace} narrative works towards aligning actions and values to promote self-acceptance. Finally, narratives rooted in the freedom moral foundation include the \emph{duty to fight}, focusing on converting others to the vegan cause and generating global change for animal protection through protests, product boycotting, and lobbying initiatives, and the \emph{right to choose}, grounded in a philosophy of respect, inclusivity, and kindness.

\subsection{Operationalization}
The \emph{moral vegan ideology} relies on two key concepts for distinguishing narrative types: collective identity and moral foundations. We operationalize them separately, as follows.

\paragraph{Collective identity} 
To operationalize collective identity, we extracted identity markers from text, focusing on first-person pronouns in both singular and plural forms. Previous studies have demonstrated the predictive nature of these pronouns in expressing collective identity in terms of agency and communion~\cite{decter2016impressive}. We derived a \emph{Collective Identity Index} (CI Index) from pronoun fractions:
\begin{equation}
\text{CI Index} = 0.5 + \frac{0.5 \cdot (\text{\emph{f}}_{\text{I}} - \text{\emph{{f}}}_{\text{we}})}{\text{\emph{f}}_{\text{I}} + \text{\emph{{f}}}_{\text{we}} + 1}
\label{eq:collective_identity_index}
\end{equation}
where $\emph{f}_{\bullet}$ denotes the relative frequency of a pronoun group in the text. The index is centered around 0.5: values around 0.5 suggest a balance between the use of first-person singular and plural pronouns, while values closer to 0 or 1 indicate a focus on the community or the self, respectively. 
Values shifted toward the two extremes of the range denote ``communal-oriented" ($index \leq 0.4$) vs. ``agency-oriented" ($index \geq 0.6$) narratives.

\paragraph{Moral foundations}
In the realm of moral foundations, the \emph{moral vegan ideology} explicitly acknowledges three crucial factors: \emph{sanctity of life}, \emph{enacting the authentic self}, and \emph{freedom}, which play discriminative roles in defining storytelling types. The Moral Foundation Theory~\cite{graham2013moral} (MFT), encompassing pillars such as \emph{care, fairness, loyalty, authority}, and \emph{sanctity}, provides a natural foundation to quantify these factors. Notably, the concepts of \emph{sanctity} and \emph{care} within MFT directly align with \emph{sanctity of life} and \emph{enacting the authentic self} in the narrative framework.
Given the inherent relationship between the pillars of MFT and the moral foundations identified in the narrative framework, we opted to leverage MFT as a means to operationalize the latter. While several dictionary-based tools exist for extracting MFT dimensions from text --- such as the extensive lexicon developed by \cite{araque2020moralstrength}, which demonstrated good performance on multiple Twitter datasets --- they have limitations. Despite their popularity and interpretability, these tools are constrained by fixed and limited vocabularies, only partial handling of word variations, and a tendency for longer documents to have a higher chance of dictionary matches. 
In response to these limitations,~\citet{nguyen2024measuring} introduced \texttt{mformer}, a transformer-based tool for extracting MFT dimensions from text. This tool outputs a number in a range from 0 to 1 for each moral dimension. It has shown superior performance compared to dictionary-based, embedding-based, and supervised classification methods. We employed \texttt{mformer} as our reference computational model for extracting moral foundations from MFT. To address potential biases in interpreting standalone moral foundation scores, we implemented a scaling process using a baseline dataset.

\begin{figure*}[ht!]
    \centering   
    \includegraphics[width=0.7\textwidth]{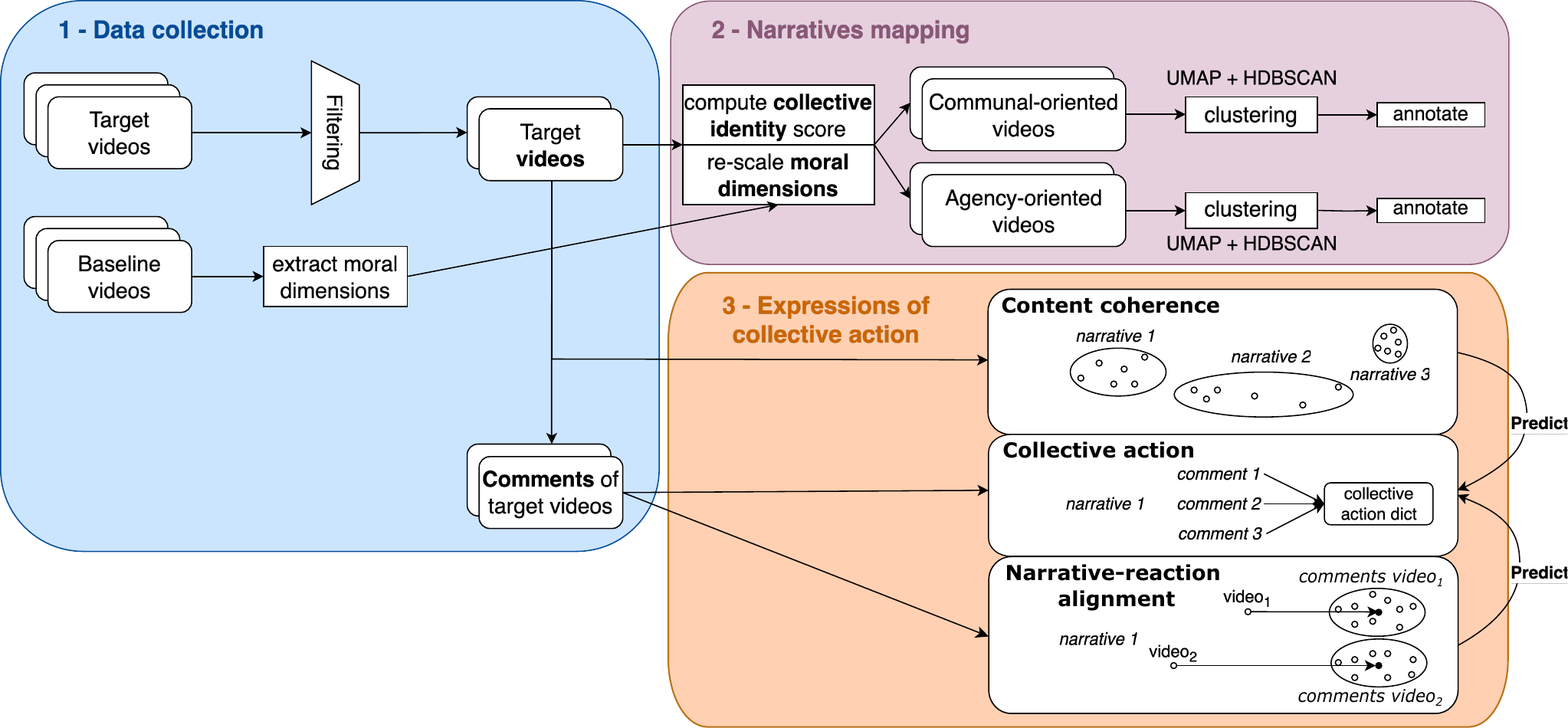}
    \caption{Methodological framework.}
    \label{fig:workflow}
\end{figure*}

\section{Methodological framework}

We aim to explore the interplay between narrative types in vegan-related videos and the levels of user engagement in collective action shown in the comments corresponding to these videos. To do so, we devised a three-step computational framework, illustrated in Figure~\ref{fig:workflow}. 
First, we collected and processed YouTube videos and associated comments. Second, we mapped video content to narrative types based on the \emph{moral vegan ideology} framework, thus obtaining distinct narrative clusters. Third and last, we associated these narrative clusters with audience reactions, specifically by investigating the presence of linguistic markers of collective action within video comments. To better interpret the relationship between the narrative clusters and their corresponding comments, we measured how much narratives are clustered in the semantic space, and we analyzed the alignment between the semantics of the video content and that of the corresponding comments.

\subsection{Data collection}\label{subsec:data_collect}
We used the YouTube API to analyze discussions related to three plant-based challenges gathering about 800k officially registered participants worldwide: \emph{Veganuary}, \emph{Meatless March}, and \emph{No Meat May}. Our selection criteria were driven by the widespread participation and popularity of \emph{Veganuary}, the leading challenge in this domain. In addition, we considered \emph{Meatless March}, and \emph{No Meat May} as they are analogous time-constrained initiatives that enjoy a substantial level of public interest.
Our data retrieval spanned videos posted on the platform during the three months centered on the reference month for each challenge, covering a 10-year period from December 2013 to June 2023.
We started the collection process by considering the lower-cased names of the movements as a primary keyword (i.e., \emph{veganuary, meatless march, no meat may}) and performed an expansion taking the relevant hashtags that were most often co-occurring with them. In practice, we first queried the API using the seed keywords, extracted the ten most frequent keywords in the descriptions of the set of retrieved videos, and manually selected those that were semantically relevant to the challenge. The final sets of keywords are in Table~\ref{tab:keywords}.

\begin{table}[t!]
\centering
\begin{tabular}{c|l}
                                                                   & \multicolumn{1}{c}{\textbf{Keywords}}                                                                                             \\ \hline
\textbf{Veganuary}                                                  & \textit{\begin{tabular}[c]{@{}l@{}}veganuary, \#veganuary, \#veganuaryYEAR, \\ \#Veganuary, \#vegan, \#Vegan, vegan\end{tabular}} \\ \hline
\textbf{\begin{tabular}[c]{@{}c@{}}Meatless \\ March\end{tabular}} & \textit{\begin{tabular}[c]{@{}l@{}}meatless march, \#meatlessmarch, \\ \#MeatlessMarch, meatless\end{tabular}}                    \\ \hline
\textbf{\begin{tabular}[c]{@{}c@{}}No Meat \\ May\end{tabular}}    & \textit{\begin{tabular}[c]{@{}l@{}}no meat may, \#nomeatmay, \#NoMeatMay, \\ vegan, \#vegan, \#Vegan\end{tabular}}               
\end{tabular}
\caption{Keywords used for the data collection by initiative.}
\label{tab:keywords}
\end{table}

We retrieved a total of 12,753 contributions with more than 5M comments. Appendix A provides a yearly breakdown of the data volume for the data of each challenge.

To analyze video content, we leveraged video transcripts. We filtered the set of videos to retain only English content, achieved through the \texttt{langdetect} tool for language detection transposed from Java in Python~\cite{shuyo2010language}. 
A subset of videos had the auto-caption option enabled by their creators and thus we were able to retrieve their transcript through the \texttt{Youtube Transcript} package~\cite{depoix2023youtube}. To complete the dataset, we employed the \texttt{Whisper} audio-to-text converter to obtain the missing data~\cite{radford2023robust}. This tool has demonstrated performance comparable to humans in terms of Word Error Rate in long-form transcription tasks.

To ensure the selection of pertinent content that could effectively be mapped to theory-defined narratives and evoke specific collective reactions, we employed Latent Dirichlet Allocation (LDA)~\cite{blei2003latent} for topic modeling and conducted a manual inspection of the results. To weed out irrelevant content from the rest, we ran LDA with two as the number of topics, we identified the ten most common words per topic and examined the top-5 most representative video transcripts for each topic. We observed that one topic predominantly revolved around recipes and ingredients, showing no relevance to the \emph{moral vegan ideology} theory, while the other delved into plant-based challenges and their associated ethical aspects. In our narrative mapping approach, the emphasis lies on achieving high precision rather than high recall. This entails prioritizing the accurate identification of pertinent elements, even at the expense of potentially overlooking some less relevant ones. As such, we filtered out recipe-focused videos (i.e. one of the retrieved topics) reaching a total volume of 3,547 data points, 1,801 of which have at least one comment with more than five unique words.

Generally speaking, our pre-processing steps were kept minimal, involving the exclusion of transcripts with fewer than five unique words and the removal of text marked with music tags indicating musical pieces within a YouTube transcript. 
Table \ref{tab:data_volume} summarizes the filtering steps and resulting data volumes.
\begin{table}[]
\begin{tabular}{c|c|c|c|c}
\textbf{Original}      & \textbf{English}       & \textbf{\begin{tabular}[c]{@{}c@{}}With \\ transcript\end{tabular}} & \textbf{\begin{tabular}[c]{@{}c@{}}Valid \\ topic\end{tabular}} & \textbf{\begin{tabular}[c]{@{}c@{}}Valid \\ comments\end{tabular}} \\ \hline
\multirow{3}{*}{12,753} & \multirow{3}{*}{11,854} & \begin{tabular}[c]{@{}c@{}}8,954\\ \emph{(YouTube)}\end{tabular}          & \multirow{3}{*}{3,547}                                              & \multirow{3}{*}{1,801}                                                   \\ \cline{3-3}
                       &                        & \begin{tabular}[c]{@{}c@{}}2,399\\ \emph{(Whisper)}\end{tabular}             &                                                                    &                                                                        
\end{tabular}
\caption{Video data volume by filtering step. Filters are applied in series, left to right.}
\label{tab:data_volume}
\end{table}

For each of the three challenges, we gathered a \emph{baseline set} of videos. These videos served us as a benchmark for re-scaling the metrics calculated on the target video set. To select the baseline, we identified the most frequent YouTube video category for each challenge. Within such categories, we then collected a set of videos of a size similar to the target set (see Appendix A) and within the same reference timeframe, without applying any keyword filter.

Last, we gathered the comments corresponding to our final set of videos. Specifically, we used the \texttt{CommentThreads} method in the YouTube API to gather up to 1000 comments and first-level replies for each video. We collected the most recent comments because, at the time of writing, the YouTube API does not allow for retrieving a random sample of comments. In total, we collected 516,207 comments on YouTube, which were further refined to 416,736 after simple pre-processing (removing mentions and URLs, and filtering out comments with fewer than five unique words).

\subsection{Narratives mapping}\label{subsec:narrative_mapping}
To map videos to narrative types defined by the \emph{moral vegan ideology}, we first grouped content by measuring the level of collective identity expressed in the transcripts (Eq.~\ref{eq:collective_identity_index}). We grouped videos into \emph{communal-oriented} (CI index $\leq 0.4$) and \emph{agency-oriented} (CI index $\geq 0.6$). We discarded the set of videos with CI index scores between 0.4 and 0.6, as they could not be strongly characterized according to this theoretical dimension. 
We applied \texttt{mformer} to the remaining 3,045 videos to extract their moral foundation scores. To find clusters of narrative types, we used UMAP~\cite{mcinnes2018umap} to reduce the scores to a 2-dimensional space, and finally applied the HDBSCAN clustering algorithm~\cite{campello2013density}.

We defined optimal HDBSCAN parameters by optimizing the Density Based Cluster Validity (DBCV)~\cite{moulavi2014density} through a random search: we set $\texttt{min\_sample}=15$, $\texttt{min\_cluster\_size}=15$, and $\texttt{metric}=\text{manhattan}$ for the communal-oriented group of videos and $\texttt{min\_sample}=15$, $\texttt{min\_cluster\_size}=150$, and $\texttt{metric}=\text{euclidean}$ for agency-oriented one.

We employed manual annotation to validate the resulting video content clusters. Acknowledging the utility of open-source models like Llama-2~\cite{touvron2023llama2} in aiding human annotators in the classification of social science concepts~\cite{ziems2023can}, we considered Llama-2 70B Chat as an initial supporting tool. For each cluster in the communal-oriented and agency-oriented groups, we selected the top 30 videos closest to the cluster centroid and prompted the language model to label each transcript within this subset according to the most relevant narrative type, if discernible. We defined narrative types for each identity group in the prompt and instructed it to output ``other'' in cases of uncertainty (refer to Appendix B for prompt descriptions). Notably, we observed a tendency of the LLM to predominantly output a single label per collective identity group (e.g., ``duty to educate'' in the communal-oriented group and ``right to good health'' in the agency-oriented one), likely influenced by its internal knowledge bias.
Given such a first initial guideline, we engaged in manual inspection of transcripts within the top-30 videos subset. This step allowed us to label each cluster and qualitatively define the theme characterizing each narrative group based on sub-categories extracted from the theory, described in Table~\ref{tab:narratives_categories}. Data volume for each annotated narrative group can be found in Appendix A.

\subsection{Collective action expressions}

We identified expressions promoting collective action within YouTube comments, aiming to quantify the relationship between these linguistic indicators and various narrative types. This analysis was inspired by theories of normative alignment, which propose that narratives advocating social and political action are crucial in igniting and sustaining collective initiatives~\cite{thomas2009aligning}. 

To further enrich our understanding of collective action markers, we incorporated two additional semantic similarity measures that previous research has associated with collective action. First, we measured the semantic coherence of videos within a narrative group, as consistent narratives can better foster a shared identity among community members, facilitating collective action efforts~\cite{blair2003exit, van2008toward}. Second, we measured the semantic alignment between the content of a video and its corresponding comments as a proxy for agreement between influential leaders and activists, which is typically associated with successful collective movements~\cite{thomas2009aligning}. Last, we fit an Ordinary Least Squares (OLS) regression to determine whether the variability in collective action responses to videos is significantly associated with the moral features emerging from them, the semantic coherence with videos' reference narratives, and the video-comments alignment.

\paragraph{Operationalization of collective action}
Aiming at the definition of a standalone measure of collective action on textual traces,~\citet{smith2018after} formulated and validated a \emph{collective action dictionary} comprising 47 terms, designed in a LIWC-like fashion~\cite{LIWC2015}. We used this dictionary to extract collective action instances through a wildcard match of words in comments. We then computed the relative frequency of collective action words in each comment and evaluated the distribution within narrative types.

\begin{figure*}[ht!]
  \centering
  \subfigure[Communal-oriented.]{\includegraphics[width=0.45\textwidth]{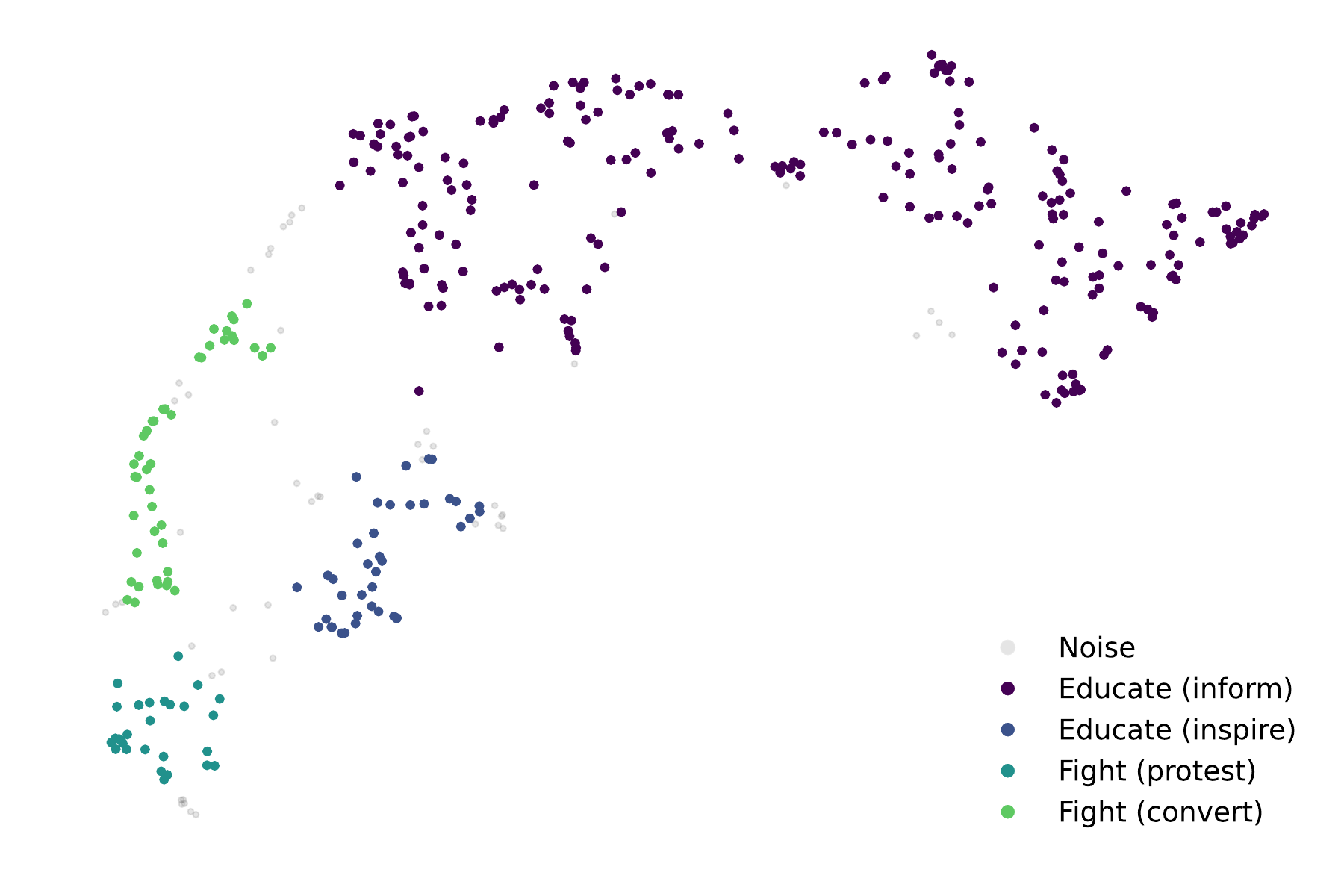}}
  \hspace{0.2em}
  \subfigure[Agency-oriented.]{\includegraphics[width=0.45\textwidth]{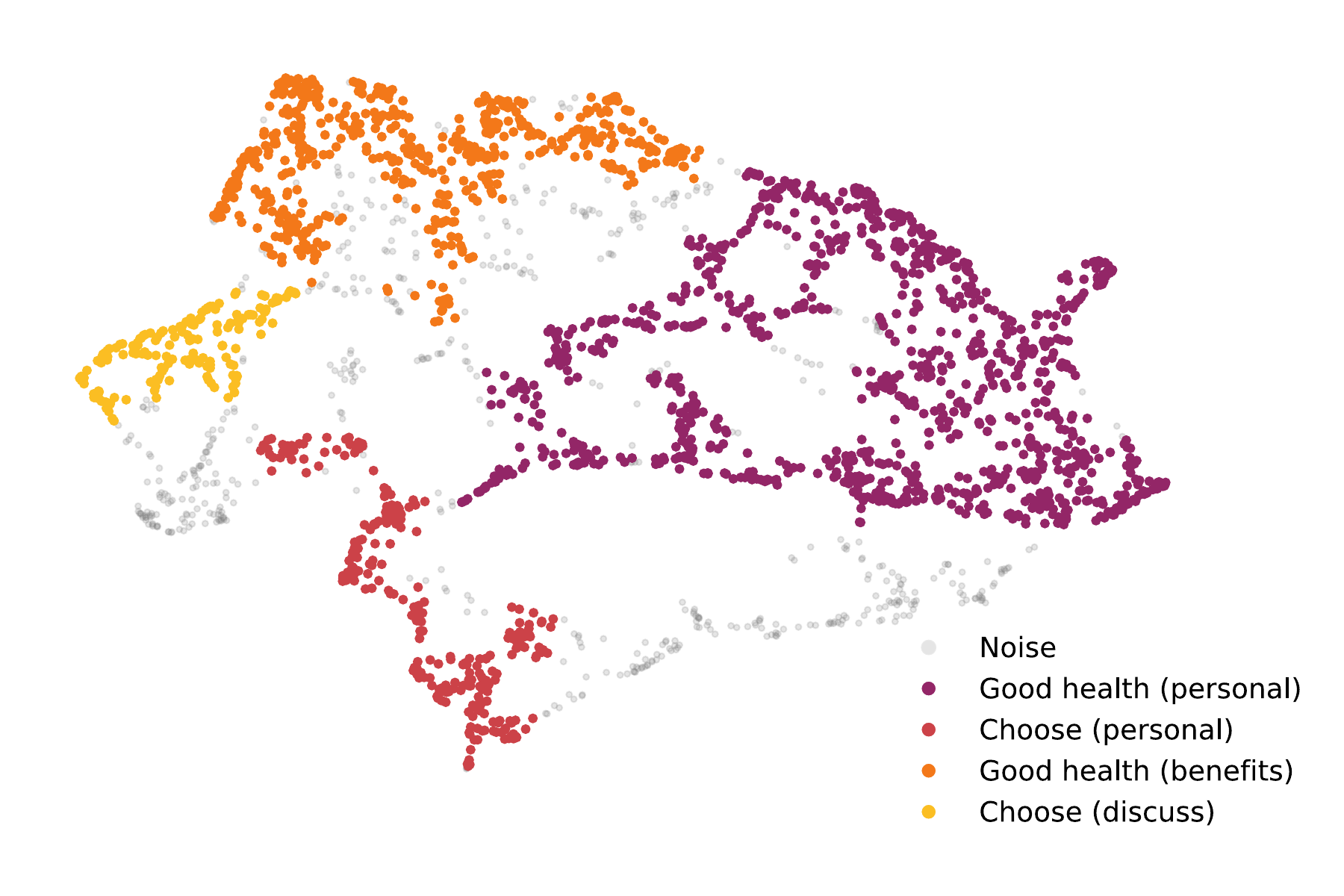}}
  \caption{Clusters of videos and their mapping to the \emph{moral vegan ideology} framework.}
  \label{fig:narratives_clusters}
\end{figure*}

\begin{table*}[ht!]
\centering
\begin{tabular}{c|l}
\textbf{Narrative}                                              & \textbf{Transcript excerpts} \\ \hline
\begin{tabular}[c]{@{}c@{}}Educate\\ (inform)\end{tabular}                                                      & \begin{tabular}[c]{@{}l@{}}Spanish startup creates 3D printed vegan meat. An Italian bioengineer at a Spanish startup\\claims to have invented the world's first vegan steak made using 3D printing technology...   \end{tabular}                       \\ \hline
\begin{tabular}[c]{@{}c@{}}Educate\\ (inspire)\end{tabular}                                                           &   \begin{tabular}[c]{@{}l@{}}I'm Matthew Glover, I'm the co-founder of Veganuary... we started it in 2013 we've had\\four campaigns now. In the first year January 2014 we have 3,300 people taking part and\\we've been doubling in size or slightly more...\end{tabular}         \\ \hline
\begin{tabular}[c]{@{}c@{}}Fight\\ (protest)\end{tabular}                                                            & \begin{tabular}[c]{@{}l@{}}...which is why we started this campaign a couple of months ago so um this is an example\\of a protest we're doing, all of our protests are not quite as destructive as this one...\end{tabular}         \\ \hline
\begin{tabular}[c]{@{}c@{}}Fight\\ (convert)\end{tabular}                                                             &     \begin{tabular}[c]{@{}l@{}}In the afternoon of August 26th 2011, we decided to
stop paying other people to kill\\animals for us to eat... We decided to stop exploiting animals for clothing, entertainment,\\testing, and all other purposes...\end{tabular}            \\ \hline
\begin{tabular}[c]{@{}c@{}}Good health\\(personal stories)\end{tabular}  &        \begin{tabular}[c]{@{}l@{}}Being a vegan bodybuilder is really tough. If you're one or want to become one, my hat is\\off to you because it's really tough...\end{tabular}          \\ \hline
\begin{tabular}[c]{@{}c@{}}Choose\\(personal stories)\end{tabular} &        \begin{tabular}[c]{@{}l@{}}I'm not a vegan. Look, man, I'm really proud of you. Cause going vegan is not something\\that I could do. You make it seem so simple as if I could choose...\end{tabular}             \\ \hline
\begin{tabular}[c]{@{}c@{}}Good health\\(benefits)\end{tabular} &       \begin{tabular}[c]{@{}l@{}}Hey this is Ryan of happy healthy vegan. So I just got back results for my latest blood\\test and I'm gonna share with you right now my testosterone levels...\end{tabular}           \\ \hline
\begin{tabular}[c]{@{}c@{}}Choose\\(discuss)\end{tabular}                                                       &    \begin{tabular}[c]{@{}l@{}}If you've ever had a
conversation with a vegan, especially if the vegan is someone you're\\close to, you know that things can get heated pretty quickly...\end{tabular}       
\end{tabular}
\caption{Representative examples of YouTube video transcripts for each narrative cluster.}
\label{tab:narrative_example}
\end{table*}

\paragraph{Content coherence} To gauge the semantic consistency of a narrative type, we evaluated the \emph{silhouette score} for each video within it, using S-BERT embeddings~\cite{reimers2019sentence} (\texttt{all-MiniLM-L6-v2} model) of video transcripts. The embedding model, trained on sentence-level data, demonstrates strong performance across various natural language processing downstream tasks. Nonetheless, its effectiveness may fluctuate when handling very large documents, often due to truncation issues.
The silhouette score is a widely-used metric to evaluate cluster quality, and measures the similarity of a datapoint with others in its own cluster compared to the similarity with points in other clusters. The distribution of silhouette scores across videos associated with a given narrative type assesses the content coherence of such videos in terms of their placement in the specific narrative cluster. 

\paragraph{Narrative-reaction alignment}
Although our study is solely observational and does not definitively establish a causal link between video content and public reaction, the nature of video-sharing platforms, which are centered around content, provides creators with a degree of influence over emerging topics in public discourse. Thus, we can study the ability of communicators to steer the audience's opinion by analyzing the relationship between published content and related reactions. This relationship can provide insights into the emergence of collective action initiatives within the reactions. We considered a \emph{video-comment alignment} metric based on the cosine similarity of pairs of videos and comments S-BERT embeddings. Given $C_i = (c_{i1}, c_{i2}, ..., c_{il})$ comments to video $v_i$, $\textbf{v}_i$ S-BERT vector of $v_i$, and $\overline{\textbf{C}_i}$ average S-BERT vector of $C_i$, the metric can be defined as:
\begin{equation}
    \text{\emph{video-comment alignment}}(v_i) = cos({\overline{\textbf{C}_i}}, \textbf{v}_i)
\label{eq:video-comment-alignment}
\end{equation}

\section{Results}

\subsection{Narratives}

\begin{figure}[t!]
    \centering   
    \includegraphics[width=0.48\textwidth]{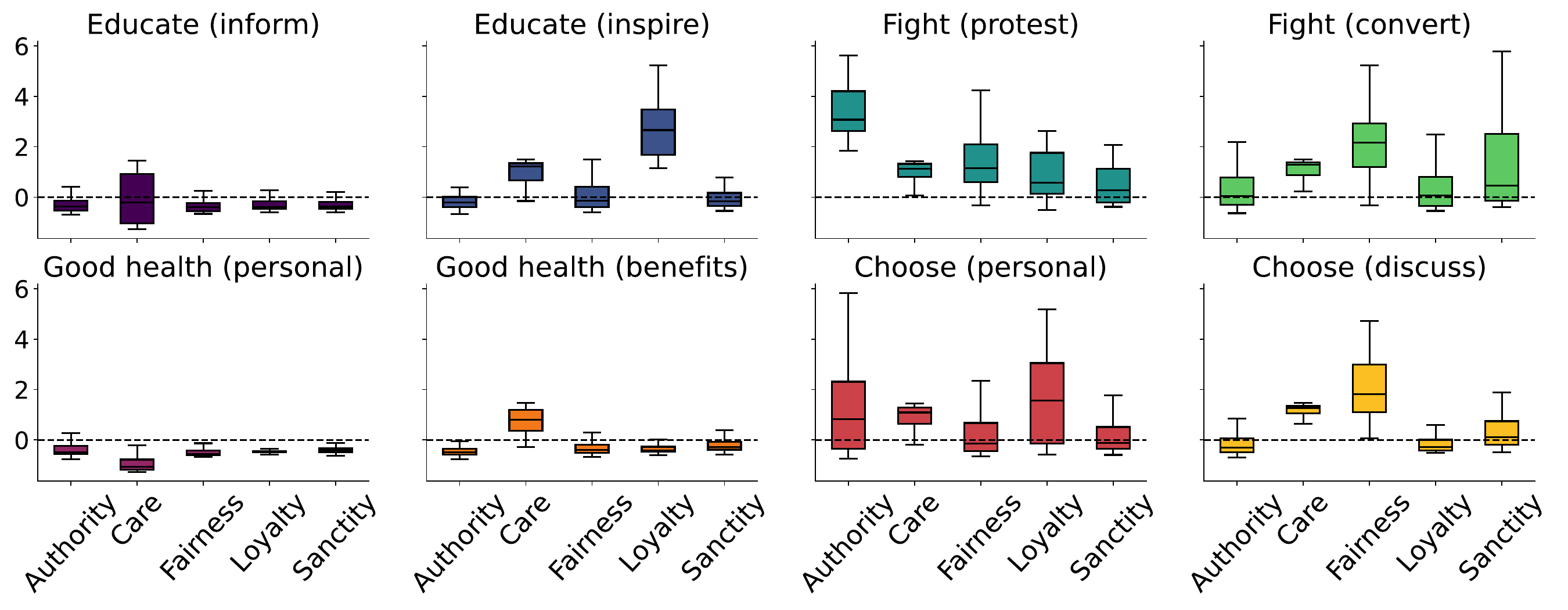}
    \caption{Moral dimensions scores for communal-oriented narratives (top) and agency-oriented narratives (bottom). Scores are discounted by the scores calculated on a control set of videos, and then standardized.}
    \label{fig:boxplot_mft}
\end{figure}

Using the \emph{Collective Identity} index (Eq.~\ref{eq:collective_identity_index}), we identified 389 communal and 2,656 agency-oriented videos. 
Within each group, the clustering on the UMAP-reduced dimensions of the Moral Foundations Theory (MFT) resulted in four clusters, as shown in Figure~\ref{fig:narratives_clusters}. We aligned each cluster with theory-defined narrative types through manual assessment. Table~\ref{tab:narrative_example} summarizes the labeling of narratives and provides a video transcript example for each of them. We also characterized each narrative based on the distribution of MFT scores of their respective videos (Figure~\ref{fig:boxplot_mft}).

Within the communal-oriented narratives, two clusters were associated with the \emph{duty to educate} narrative. The first primarily focuses on information sharing, embodying the creator's role as a ``conveyor of knowledge" (\emph{educate (inform)}). Videos within this cluster exhibit a marked character of news-like dissemination, therefore their transcripts do not generally show high levels in any MFT dimension. The second cluster emphasizes the inspirational aspect of storytelling, motivating others to start their journey toward veganism or persist in their pursuit (\emph{educate (inspire)}).
Its videos exhibit higher levels of loyalty compared to other clusters, reflecting the feeling of ``one for all and all for one" typical of that narrative type. The remaining two clusters were linked to the \emph{duty to fight} narrative. The first expresses the intent to persuade others to act towards broader global change (\emph{fight (convert)}), and it is characterized by a high level of the \emph{fairness} MFT dimension, which is linked to the evolutionary process of reciprocal altruism~\cite{brosnan2002proximate}. The latter promotes active participation in protests and lobbying to attain climate objectives (\emph{fight (protest)}), and does so by expressing high levels of the \emph{authority} dimension, which stresses the values of leadership and followership.

Within agency-oriented narratives, two clusters refer to concepts typical of \emph{right-to-good-health} narratives: emphasizing personal experiences and transformations (\emph{good health (personal stories)}) and praising the physical and mental health benefits of plant-based diets (\emph{good health (benefits)}). The MFT dimensions for these two clusters are not particularly distinctive, with scores for most dimensions being at the baseline level. The remaining two clusters are semantically associated with the \emph{right-to-choose} narrative, and address the need for broadening the societal acceptance of veganism. The first cluster does so by sharing personal stories (\emph{choose (personal stories)}), which exhibit on average high scores of the \emph{loyalty} dimension being linked to the moral's defining sense of self-sacrifice. The second cluster encourages instead informed discussions (\emph{choose (discuss)}) and the language used in its videos is high on the \emph{fairness} dimension, highlighting the virtues of justice and rights.

\subsection{Collective action in reactions}

\begin{figure}[t!]
    \centering   
    \includegraphics[width=0.31\textwidth]{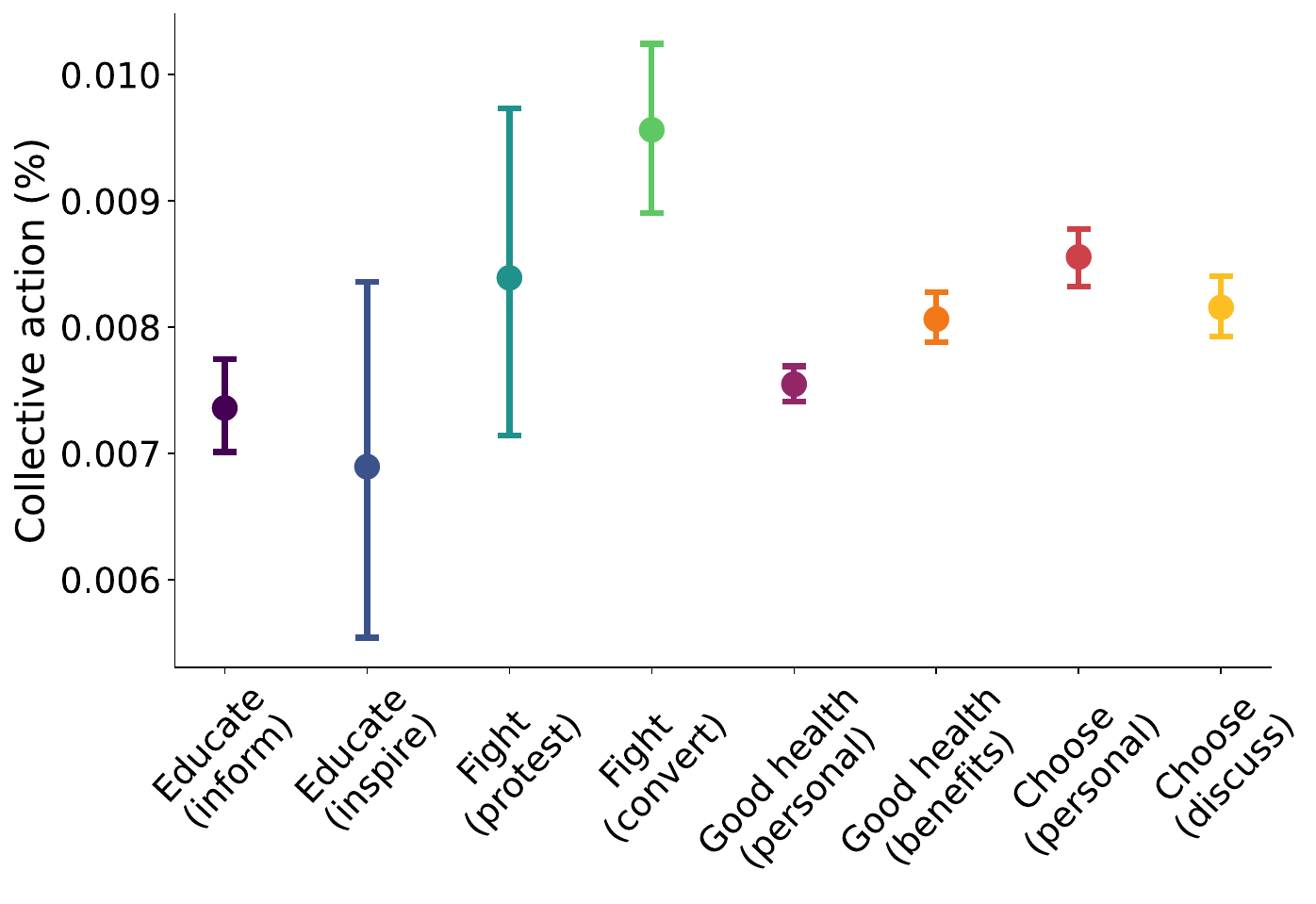}
    \caption{Relative frequency of collective action markers in comments by narrative type.}
    \label{fig:collective_action}
\end{figure}

We examined the frequency of words associated with collective action in comments on videos across narrative groups (Figure \ref{fig:collective_action}). We observed a variation in the frequency of these linguistic markers across narratives, with the highest frequency in videos that fell under the \emph{fight (convert)} narrative. Appendix C reports some examples of comments characterized by high levels of collective action.

\begin{figure}[t!]
  \centering
  \subfigure[Silhouette score.]{\includegraphics[width=0.23\textwidth]{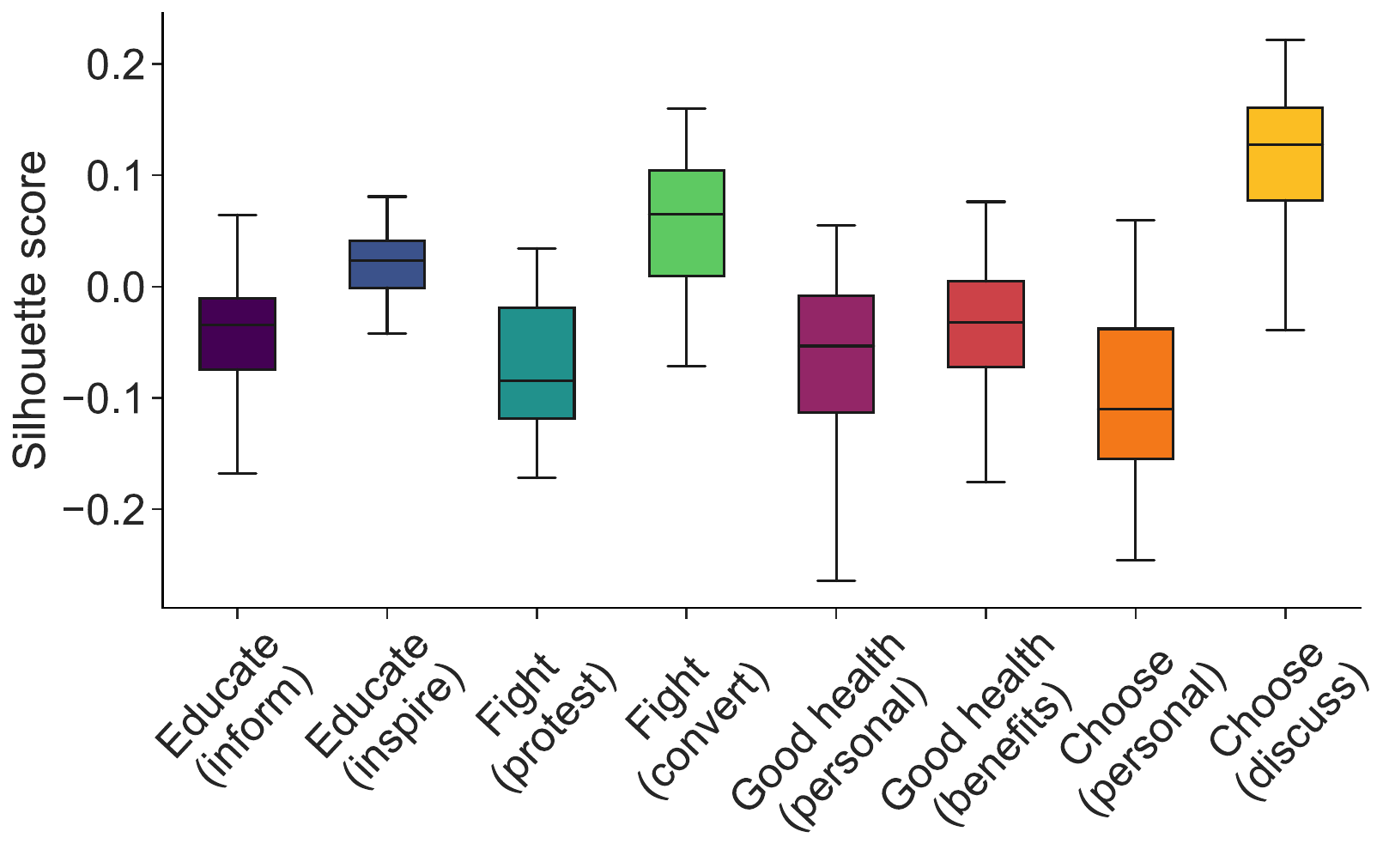}}
  \hspace{0.005em}
  \subfigure[Video-comment align.]{\includegraphics[width=0.23\textwidth]{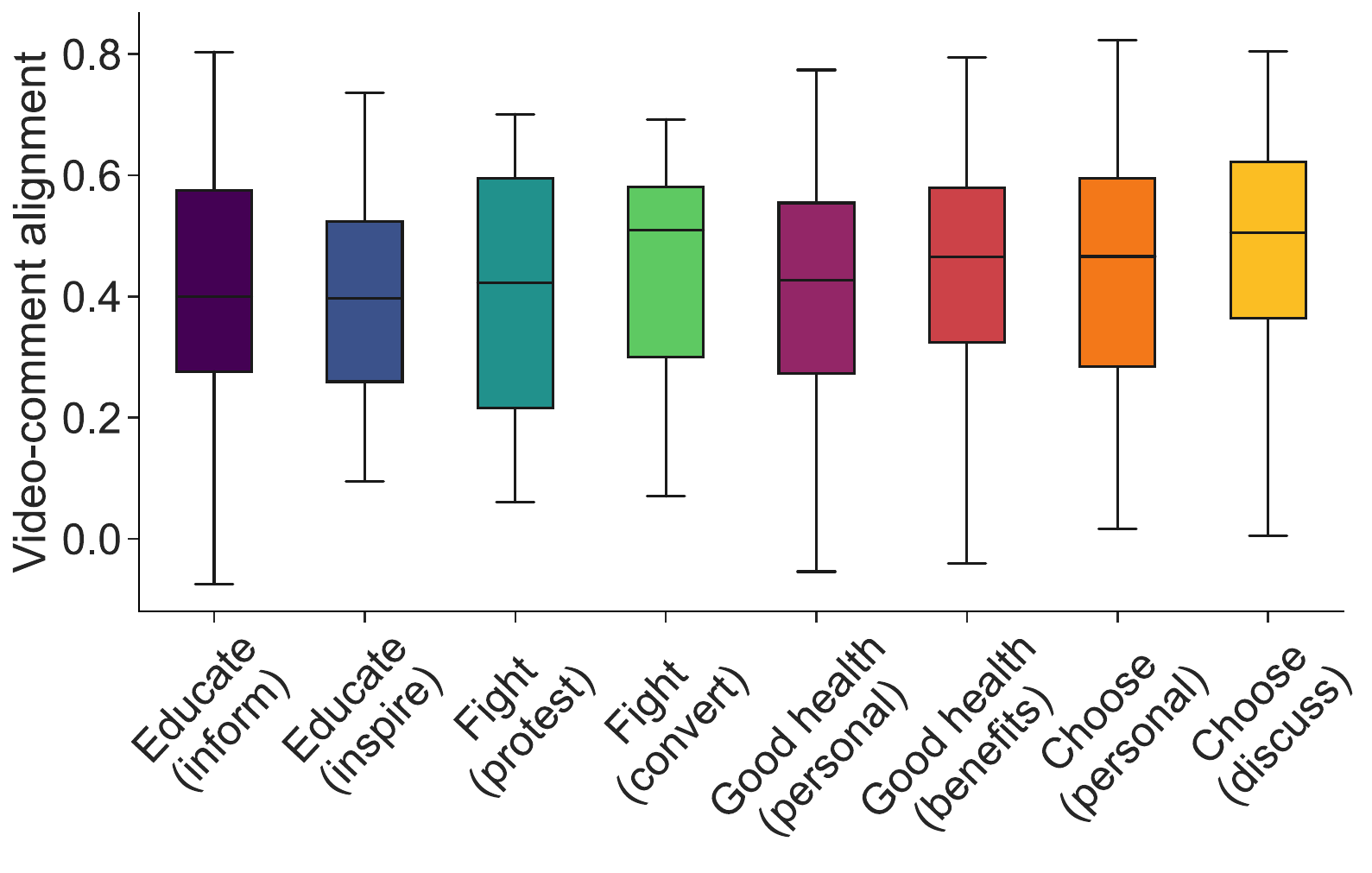}}
  \caption{Content coherence and narrative-reaction alignment by narrative type.}
  \label{fig:boxplots_semantic}
\end{figure}

\begin{figure}[t!]
  \centering
  \subfigure[Fight (convert).]{\includegraphics[width=0.48\columnwidth]{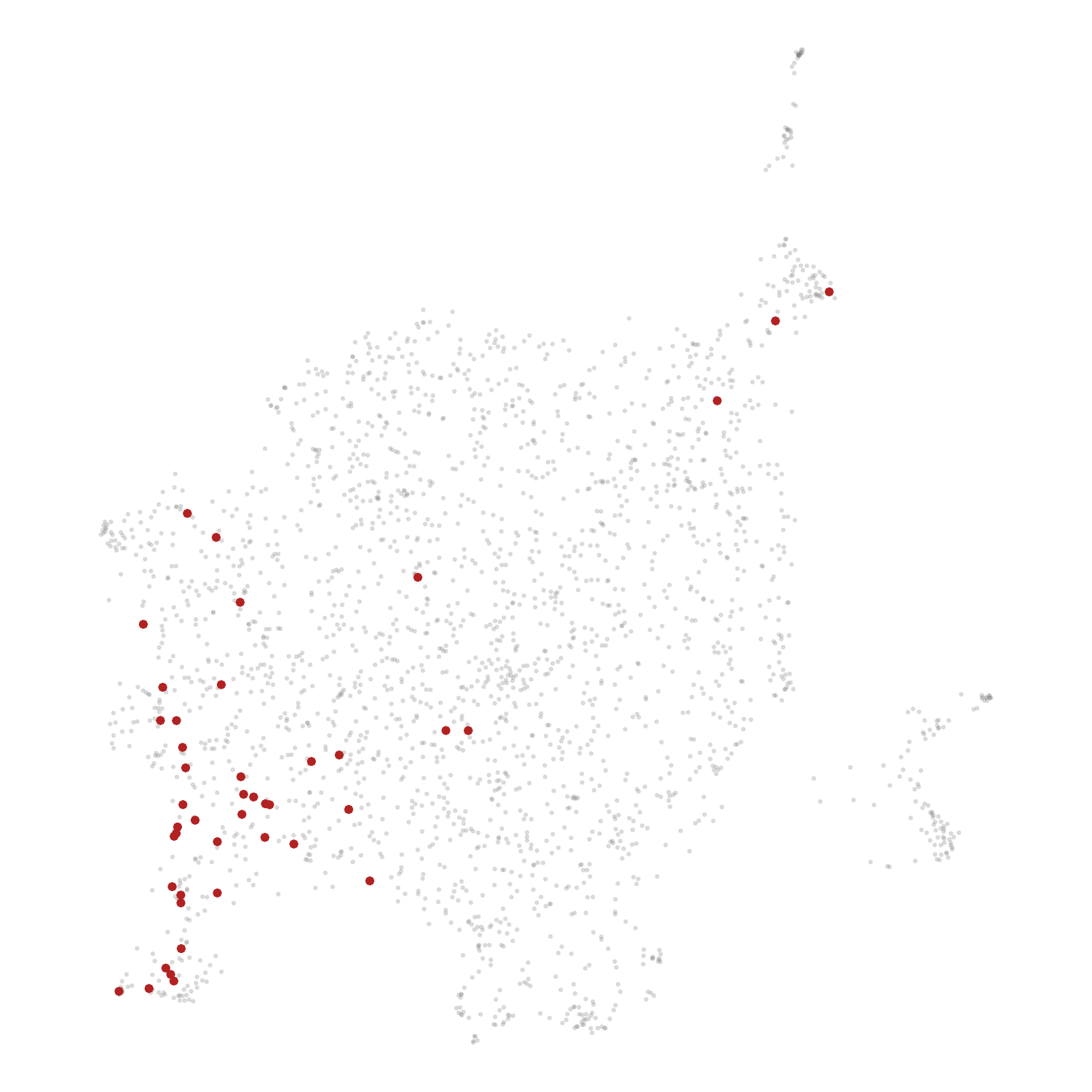}}
  \hspace{0.2em}
  \subfigure[Choose (discuss).]{\includegraphics[width=0.48\columnwidth]{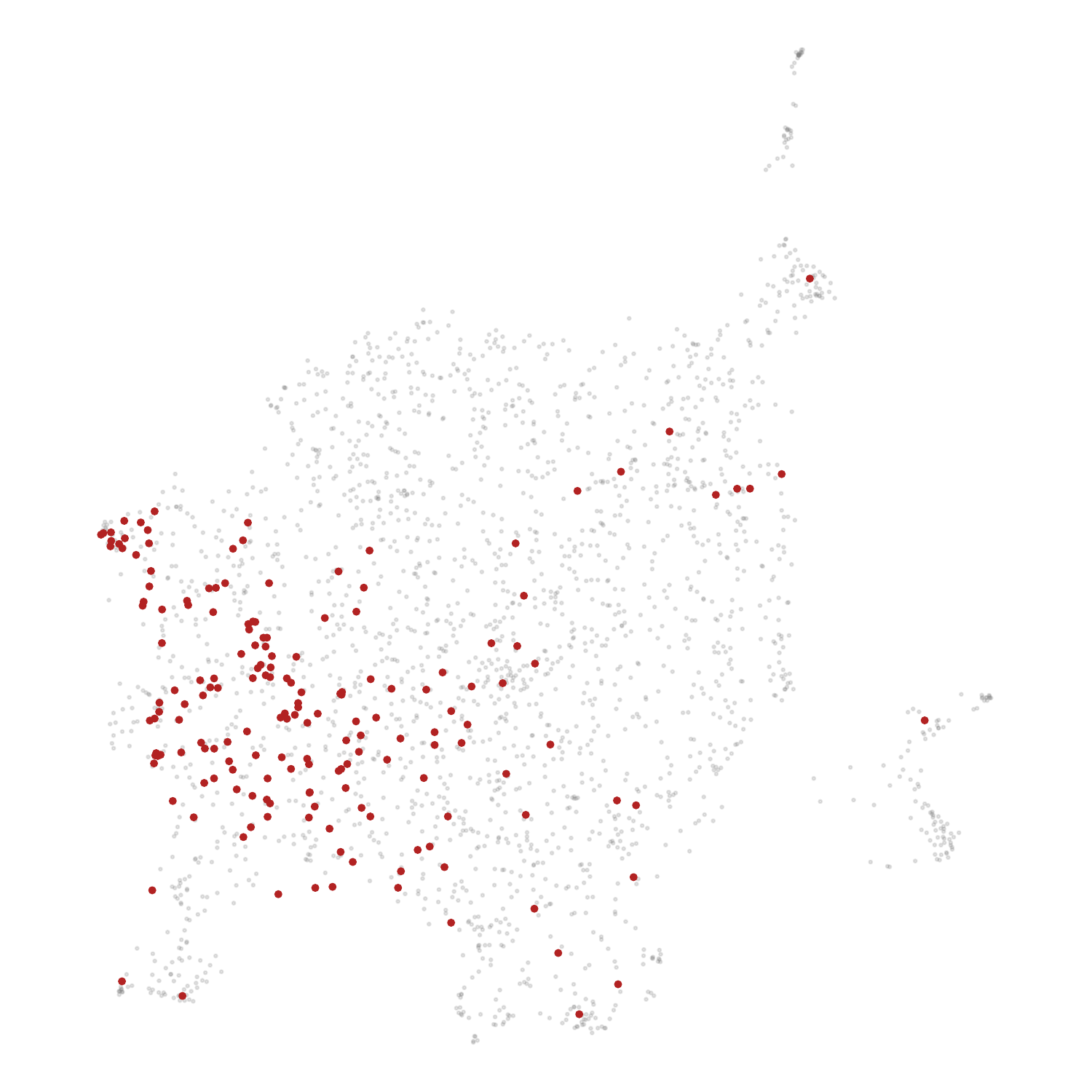}}
  \caption{UMAP projections of S-BERT video embeddings of all videos, with videos belonging to two narrative types highlighted in red.}
  \label{fig:video_umap}
\end{figure}

We further investigated the relationship between the frequency of commitment markers, video content coherence within narrative groups, and narrative-reaction alignment (Eq.~\ref{eq:video-comment-alignment}). The narratives within the \emph{fight (convert)} and \emph{choose (discuss)} clusters demonstrated the highest overall levels of semantic consistency (Figure~\ref{fig:boxplots_semantic}, left). These two narratives were found to occupy a narrow semantic area in the embedding space, as shown by the UMAP projections of the S-BERT embeddings of the video transcripts (Figure~\ref{fig:video_umap}). We assessed whether the high semantic coherence of certain narrative types translated into a strong semantic alignment between content and reactions (Figure~\ref{fig:boxplots_semantic}). The \emph{fight (convert)} and \emph{choose (discuss)} narratives were characterized by a slightly higher similarity between videos and corresponding comments compared to other clusters. Overall, our findings suggest a correlation between content semantic consistency and the capacity of some narratives to attract reactions aligned with video content.

The observed macro-level association between collective action indicators, narrative types, semantic coherence of content, and alignment of comments to videos, raises the question of whether the frequency of collective action mentions in comments on individual videos can be inferred from these factors. To address this, we employed an Ordinary Least Squares (OLS) model, using individual videos as the unit of analysis, to predict the degree of commitment based on moral foundations scores, semantic coherence between videos and narrative groups, and video-comment alignment. We also factored in the number of comments on each video as a control variable to account for the video's popularity and the content creator's influence. The model was applied after adjusting skewed predictors and calculating the z-scores of all variables, as summarized in Table~\ref{tab:ols}. Despite the inherent challenge in predicting comment commitment variability at the individual video level, we identified several regression coefficients of statistical significance (p-values $\leq 0.1$). Specifically, a smaller audience size, higher semantic coherence between content and narrative type, and a discourse more focused on group self-sacrifice (key characteristics of the loyalty moral dimension) were significantly correlated with increased collective action in responses.

\begin{table}[t!]
\centering
\begin{tabular}{ccc}
\multicolumn{1}{c|}{\textbf{Variable}}                                                                 & \multicolumn{1}{c|}{\textbf{Coefficient}} & \textbf{p-value} \\ \hline
\multicolumn{1}{c|}{\textit{const}}                                                                    & \multicolumn{1}{c|}{1.16e-17}              & 1.000            \\ \hline
\multicolumn{1}{c|}{\textit{n. comments (log)}}                                                        & \multicolumn{1}{c|}{\textbf{-0.059}}      & \textbf{0.025**}   \\ \hline
\multicolumn{1}{c|}{\textit{\begin{tabular}[c]{@{}c@{}}video-comment \\ alignment (log)\end{tabular}}} & \multicolumn{1}{c|}{0.015}                & 0.576            \\ \hline
\multicolumn{1}{c|}{\textit{\begin{tabular}[c]{@{}c@{}}silhouette (log)\end{tabular}}}   & \multicolumn{1}{c|}{\textbf{0.047}}       & \textbf{0.066*}   \\ \hline
\multicolumn{1}{c|}{\textit{care score}}                                                               & \multicolumn{1}{c|}{-0.034}               & 0.292            \\ \hline
\multicolumn{1}{c|}{\textit{fairness score (sqrt)}}                                                    & \multicolumn{1}{c|}{0.024}                & 0.416            \\ \hline
\multicolumn{1}{c|}{\textit{loyalty score (sqrt)}}                                                     & \multicolumn{1}{c|}{\textbf{0.051}}       & \textbf{0.053*}   \\ \hline
\multicolumn{1}{c|}{\textit{authority score (sqrt)}}                                                   & \multicolumn{1}{c|}{-0.027}               & 0.329            \\ \hline
\multicolumn{1}{c|}{\textit{sanctity score (sqrt)}}                                                    & \multicolumn{1}{c|}{0.025}                & 0.428            \\ \hline
\multicolumn{1}{c}{\textbf{R}$^2$ = $0.008$}                                                      &                                       &                 
\end{tabular}
\caption{OLS regression coefficients to predict the fraction of comments containing linguistic markers of collective action.}
\label{tab:ols}
\end{table}

\section{Discussion and conclusion}
In pursuit of evaluating the influence of online storytelling on collective action initiatives, specifically within the context of climate action, we analyzed YouTube content related to three popular challenges aimed at promoting awareness about plant-based diets: \emph{Veganuary}, \emph{Meatless March}, and \emph{No Meat May}.

We developed the first operationalization of a theoretical framework that formalizes six moral narratives within the vegan ideology, and mapped YouTube videos to these storytelling types (\textbf{RQ1}). For communal-oriented narratives, a clear mapping emerged for the theoretical concepts of ``duty to educate" and ``duty to fight", while ``duty to care" was less distinct in the reference videos. The ``duty to educate" narrative frequently surfaced during annotation, likely because of the news-like communication tone of many YouTube vlogs. The distinction between agency-oriented narrative types was more nuanced. The narrative of the ``right to good health" was often present, even blending with other narrative types, suggesting a tendency to emphasize the benefits of plant-based diets. The ``right to choose" narrative also appeared frequently, while the ``right to inner peace" narrative was less common.

Building on the relationship between public narratives and collective action explored in previous theoretical literature, we investigated the connection between narrative types and linguistic markers indicative of collective action in video comments. Narratives advocating for a duty to fight and convert others to the veganism cause elicited the highest level of collective action responses (\textbf{RQ2}). Overall, we found that a smaller audience, greater content coherence of videos with their narrative group, and discourse expressing higher levels of the ``loyalty'' moral foundation were significantly associated with increased collective action (\textbf{RQ3}). 
These findings align with established research in the social sciences. The seminal work of~\citet{olson1971logic} posited an inverse relationship between the probability of collective action and group size. Moreover, the coherence of content within its designated narrative can cultivate a shared identity among community members, thus fostering more effective collective action efforts~\cite{blair2003exit, van2008toward}. Additionally,~\cite{willer2009groups} emphasize the intricate relationship between collective action, the benefits accrued within groups (such as status), and the notion of self-sacrifice for the greater collective—a fundamental aspect of the ``loyalty'' moral foundation. These insights contribute to a nuanced understanding of the dynamics that drive collective action within the context of the proposed analysis.

While our study provides novel quantitative insights into the relationship between narratives and collective action, it has limitations that open up avenues for future research. 
Primarily, the vegan narrative framework we employed is inherently tailored to its specific topic, as defined within the existing literature. Nonetheless, its structured and clear outline, rooted in collective identity and moral perspectives, presents the opportunity for expansion into other contexts involving lifestyle choices.
Other issues arise from the limited amount of data involved in the final steps of the study.
Notably, our data pre-processing pipeline retained only 24\% of the initial 12,753 videos potentially relevant to the veganism discourse. This was due to several filtering steps aimed at prioritizing precision over recall. Most notably, we excluded video-recipes and all videos not heavily characterized on the Collective Identity axis of our reference theory. Our operationalization covered only a subset of the narratives originally formulated by the theory, which could be due to these filtering steps or to the limited applicability of the theory to the specific context of YouTube. Another limitation pertains to the complex nature of the concepts characterizing the narratives and the public response they received. Despite using validated tools, the inherent ambiguity and complexity of some of the key theoretical notions we base our work on may have introduced noise in the final mapping between empirical clusters of videos and theoretical narratives. Last, the observed collective action in response to narrative content does not necessarily translate to offline participation in social movements, as social media often fosters a form of activism known as ``slacktivism"~\cite{morozov2009brave}.

Despite these constraints, our study successfully captures the existence of nuances in pro-environmental storytelling and discerns variations in the type of engagement that different narrative types receive.

\section{Related work}
Extensive exploration within social science literature has elucidated the profound impact of discourse framing on collective social processes, particularly within the development of social movements. \citet{polletta1998contending} identified three dynamics of collective action where stories play pivotal roles, including ``stories of origin", ``stories of defeat" and ``stories of victory". \citet{davis2002narrative} expanded on this by highlighting the interactive exchange process inherent in narrative frameworks, emphasizing the crucial role of audiences in shaping and responding to stories, with a focus on social identity aspects.

Environmental social movements, the main target of our research, have been often examined through the lens of narrative networks, emphasizing the connection between stories and the social networks formed in the context of environmental protection~\cite{lejano2013power}. Within this setting, climate change-specific narratives have been categorized into ``lifeboat" and ``collective" types, reflecting dichotomies between self-interest and shared purpose within communities~\cite{fiskio2012apocalypse}. The theoretical framework by~\citet{napoli2020vegan}, a key reference for our work, further enriched this collective identity-based description of environmental narratives by focusing on the vegan movement and adding nuances of morality. This perspective has been further corroborated by later experimental work~\cite{waters2022v,aavik2023going}.

While numerous contributions exist within the field of Narrative Analysis that extract narratives from textual traces using Natural Language Processing (NLP) tools~\cite{bandeli2020framework, debnath2020grounded, ranade2022computational}, the link between computational methods and narrative types discussed in social science literature is still under-explored~\cite{piper2021narrative}. Notable exceptions include the work of~\citet{vu2021social}, who extracted framing ingredients of climate change-related protest messages, operationalizing concepts such as impacts, action, and efficacy~\cite{benford2000framing} through mention counts.

Shifting the focus to the operationalization of collective action, the literature presents a richer landscape, often examined through network science lenses~\cite{lucchini2022reddit}.
Approaches involving textual analysis, particularly the use of recurrent expressions~\cite{berlin2008think, fetzer2008and}, present challenges especially when applied to social media comments.
Research has also explored machine learning tools to extract linguistic markers hinting at collective action from text~\cite{wang2017words, zhang2019casm, li2023sdgdetector}, yet the domain-specific nature of these approaches limits their general applicability. 
In the realm of dictionary-based methods for collective action measurement, studies have commonly focused on collective identity, emotions, and collective or participatory efficacy as predictors~\cite{van2008toward, van2013believing, rand2015collective, bamberg2015collective, van2019hope, hamann2020my, furlong2021social, brown2022opposing, judge2022dietary, thomas2022mobilise}. Some have aimed to define computational measures for these elements from text, often resorting to LIWC categories~\cite{gulliver2021assessing, suitner2023rise}. Within this group of works, \citet{smith2018after} introduced a collective action dictionary, defined in a LIWC-like fashion, which served as a valuable resource for our analysis.

\section{Ethical considerations}
Given that YouTube content is predominantly user-generated and largely unmoderated, the narratives we examined may contain messages that do not necessarily advocate for veganism and climate action in an ethical or respectful manner. Our study adopted a broad view of the public discourse on YouTube, without attempting to categorize content that may contain misinformation or potentially harmful expressions. We underscore the importance of addressing the potential ethical implications associated with unregulated online spaces.

We recognize the potential misuse of our framework, which could be exploited to inform strategies of misinformation and propaganda. Despite acknowledging the risks of certain aspects being leveraged to create tools and strategies with detrimental societal implications, we are dedicated to minimizing such risks. Our motivation is grounded in the belief that our work can positively contribute to addressing pressing social dilemmas, particularly those related to climate change.

The dataset associated with this paper complies with the FAIR principles~\cite{fair}, as outlined in the DataSheet (Supplementary Material). 

\section{Code and data availability}
All code used for the experiments and information about the videos examined, in terms of narrative orientations, statistics, metrics specific to this paper, reference baseline videos, and comments retrieved and analyzed is available on a GitHub repository\footnote{\url{https://github.com/ariannap13/VeganCollectiveAction}}.

\section*{Acknowledgments}
We acknowledge the support from the Carlsberg Foundation through the COCOONS project (CF21-0432). The funder had no role in study design, data collection and analysis, decision to publish, or preparation of the manuscript.

\bibliography{aaai22}

\clearpage

\section*{Appendix A}

Table \ref{tab:volume_video} summarizes the volume of collected data in terms of the number of videos and comments per year, for each of the analyzed plant-based challenges. Note that such a volume refers to the originally collected data, i.e. no filtering or cleaning is performed in this phase.

\begin{table}[ht]
\centering
\caption{Volume of originally retrieved videos and comments by movement and year: Veganuary (v), Meatless March (m), No Meat May (n).}
\label{tab:volume_video}
\begin{tabular}{l|l|l}
\textbf{Year} & \textbf{n. videos}                                                        & \textbf{n. comments}                                    \\ \hline
\textit{2014} & \begin{tabular}[c]{@{}l@{}}v: 300\\ m: 64\\ n: 314\end{tabular}                      & \begin{tabular}[c]{@{}l@{}}v: 27113\\ m: 281\\ n: 22380\end{tabular} \\ \hline
\textit{2015} & \begin{tabular}[c]{@{}l@{}}v: 431\\ m: 91\\ n: 423\end{tabular}                      & \begin{tabular}[c]{@{}l@{}}v: 61047\\ m: 2379\\ n: 143316\end{tabular} \\ \hline
\textit{2016}           & \multicolumn{1}{l|}{\begin{tabular}[c]{@{}l@{}}v: 445\\ m: 152\\ n: 457\end{tabular}} & \begin{tabular}[c]{@{}l@{}}v: 74136\\ m: 2715\\ n: 303835\end{tabular} \\ \hline
\textit{2017}        & \multicolumn{1}{l|}{\begin{tabular}[c]{@{}l@{}}v: 527\\ m: 175\\ n: 438\end{tabular}} & \begin{tabular}[c]{@{}l@{}}v: 202943\\ m: 3688\\ n: 166767\end{tabular} \\ \hline
\textit{2018}           & \multicolumn{1}{l|}{\begin{tabular}[c]{@{}l@{}}v: 678\\ m: 207\\ n: 331\end{tabular}} & \begin{tabular}[c]{@{}l@{}}v: 144711\\ m: 16217\\ n: 440456\end{tabular} \\ \hline
\textit{2019}           & \multicolumn{1}{l|}{\begin{tabular}[c]{@{}l@{}}v: 728\\ m: 240\\ n: 437\end{tabular}} & \begin{tabular}[c]{@{}l@{}}v: 146233\\ m: 11587\\ n: 424704\end{tabular} \\ \hline
\textit{2020}          & \multicolumn{1}{l|}{\begin{tabular}[c]{@{}l@{}}v: 820\\ m: 336\\ n: 442\end{tabular}} & \begin{tabular}[c]{@{}l@{}}v: 241289\\ m: 128576\\ n: 342716\end{tabular} \\ \hline
\textit{2021}          & \multicolumn{1}{l|}{\begin{tabular}[c]{@{}l@{}}v: 786\\ m: 392\\ n: 441\end{tabular}} & \begin{tabular}[c]{@{}l@{}}v: 418376\\ m: 18375\\ n: 398727\end{tabular} \\ \hline
\textit{2022}           & \multicolumn{1}{l|}{\begin{tabular}[c]{@{}l@{}}v: 787\\ m: 380\\ n: 305\end{tabular}} & \begin{tabular}[c]{@{}l@{}}v: 151607\\ m: 12584\\ n: 176818\end{tabular} \\ \hline
\textit{2023}           & \multicolumn{1}{l|}{\begin{tabular}[c]{@{}l@{}}v: 724\\ m: 385\\ n: 517\end{tabular}} & \begin{tabular}[c]{@{}l@{}}v: 460404\\ m: 37384\\ n: 484054\end{tabular} \\ \hline
\end{tabular}
\end{table}

Table \ref{tab:volume_compare} compares the volume of the cleaned and filtered collected target data with the volume of baseline videos of reference, for each of the analyzed plant-based challenges. 

\begin{table}[ht]
\centering
\caption{Volume of cleaned videos by movement and year, comparing target and baseline sets: Veganuary (v), Meatless March (m), No Meat May (n).}
\label{tab:volume_compare}
\begin{tabular}{l|l|l}
\textbf{Year} & \textbf{n. videos target}                                                        & \textbf{n. videos baseline}                                    \\ \hline
\textit{2014} & \begin{tabular}[c]{@{}l@{}}v: 78\\ m: 24\\ n: 81\end{tabular}                      & \begin{tabular}[c]{@{}l@{}}v: 76\\ m: 24\\ n: 76\end{tabular} \\ \hline
\textit{2015} & \begin{tabular}[c]{@{}l@{}}v: 133\\ m: 34\\ n: 136\end{tabular}                      & \begin{tabular}[c]{@{}l@{}}v: 129\\ m: 34\\ n: 136\end{tabular} \\ \hline
\textit{2016}          & \multicolumn{1}{l|}{\begin{tabular}[c]{@{}l@{}}v: 160\\ m: 39\\ n: 145\end{tabular}} & \begin{tabular}[c]{@{}l@{}}v: 141\\ m: 39\\ n: 143\end{tabular} \\ \hline
\textit{2017}         & \multicolumn{1}{l|}{\begin{tabular}[c]{@{}l@{}}v: 182\\ m: 43\\ n: 106\end{tabular}} & \begin{tabular}[c]{@{}l@{}}v: 173\\ m: 42\\ n: 103\end{tabular} \\ \hline
\textit{2018}         & \multicolumn{1}{l|}{\begin{tabular}[c]{@{}l@{}}v: 237\\ m: 46\\ n: 78\end{tabular}} & \begin{tabular}[c]{@{}l@{}}v: 220\\ m: 45\\ n: 77\end{tabular} \\ \hline
\textit{2019}        & \multicolumn{1}{l|}{\begin{tabular}[c]{@{}l@{}}v: 291\\ m: 82\\ n: 132\end{tabular}} & \begin{tabular}[c]{@{}l@{}}v: 233\\ m: 76\\ n: 129\end{tabular} \\ \hline
\textit{2020}      & \multicolumn{1}{l|}{\begin{tabular}[c]{@{}l@{}}v: 290\\ m: 69\\ n: 77\end{tabular}} & \begin{tabular}[c]{@{}l@{}}v: 246\\ m: 69\\ n: 77\end{tabular} \\ \hline
\textit{2021}       & \multicolumn{1}{l|}{\begin{tabular}[c]{@{}l@{}}v: 232\\ m: 71\\ n: 100\end{tabular}} & \begin{tabular}[c]{@{}l@{}}v: 219\\ m: 66\\ n: 96\end{tabular} \\ \hline
\textit{2022}       & \multicolumn{1}{l|}{\begin{tabular}[c]{@{}l@{}}v: 166\\ m: 63\\ n: 76\end{tabular}} & \begin{tabular}[c]{@{}l@{}}v: 151\\ m: 61\\ n: 75\end{tabular} \\ \hline
\textit{2023}     & \multicolumn{1}{l|}{\begin{tabular}[c]{@{}l@{}}v: 173\\ m: 56\\ n: 147\end{tabular}} & \begin{tabular}[c]{@{}l@{}}v: 125\\ m: 49\\ n: 88\end{tabular} \\ \hline
\end{tabular}
\end{table}

\begin{figure}[t!]
    \centering   
    \includegraphics[width=0.47\textwidth]{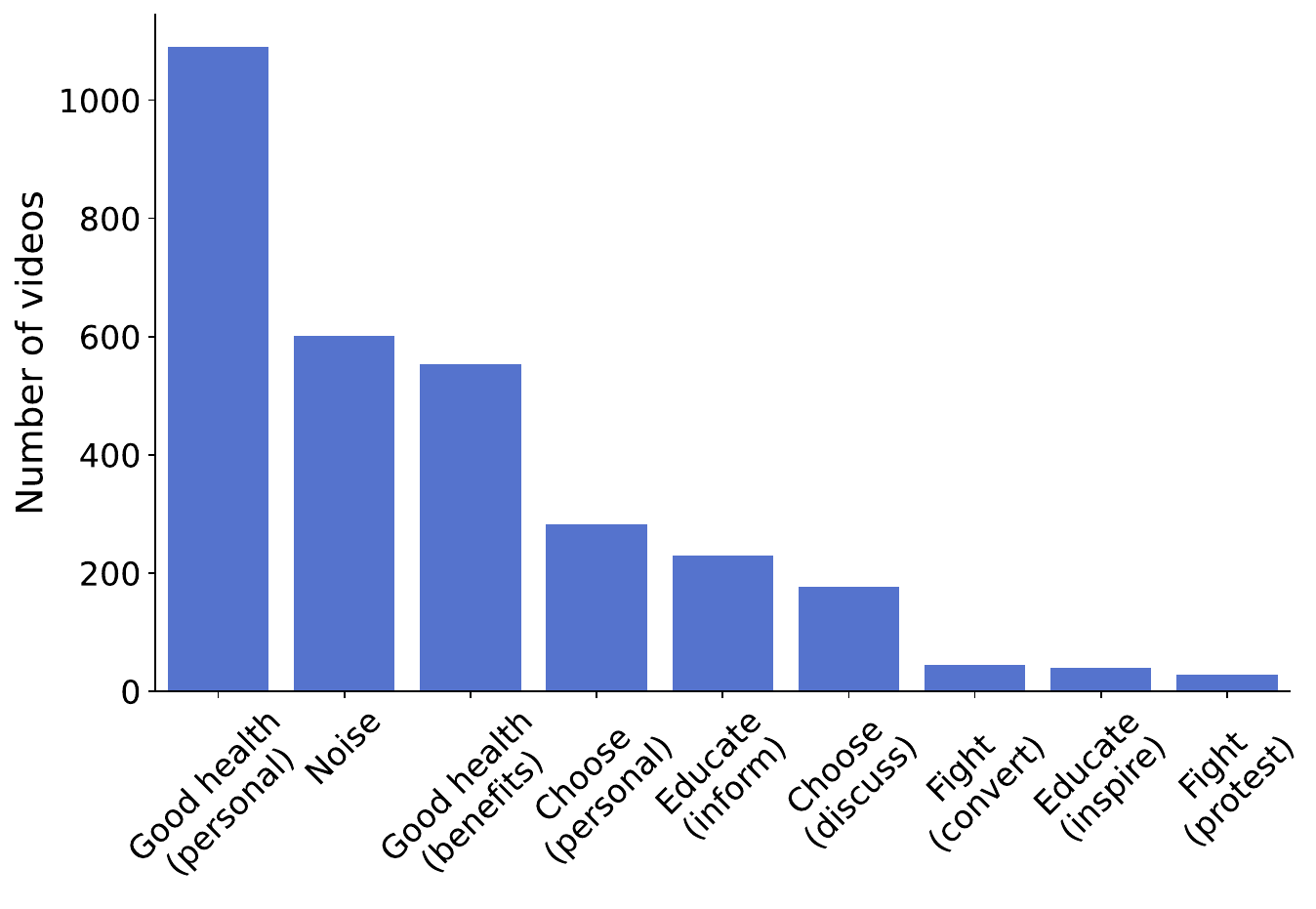}
    \caption{Number of videos by narrative type, according to the reference moral vegan ideology framework.}
    \label{fig:mapping_narrative}
\end{figure}

Figure \ref{fig:mapping_narrative} outlines the distribution of videos by mapped narrative type.

\newpage 

\section*{Appendix B}

We prompted Llama 2 70B Chat by providing the model with a system prompt to contextualize the task and a human message to precisely define the task. For each of the narrative identity groups (i.e. communal-oriented and agency-oriented), we included definitions of narratives in the prompts as follows:
\begin{itemize}
    \item \emph{Duty to care}: Moral duty to protect all living creatures, fostering compassion and connection with all beings.
    \item \emph{Duty to fight}: Moral duty to dispel ignorance, overcome misinformation, and inspire others about the benefits of a vegan lifestyle.
    \item \emph{Duty to educate}: Moral duty to actively engage in rights activism, driven by a sense of injustice.
    \item \emph{Right to good health}: Moral right to prioritize personal well-being through a plant-based diet, emphasizing physical and mental benefits.
    \item \emph{Right to inner peace}: Moral right to align actions with values, fostering a life of peace and self-acceptance.
    \item \emph{Right to choose}: Moral right to make personal choices, accepting others' opinion and not imposing own choices.
\end{itemize}

\begin{promptbox}{Narratives - communal-oriented}
System prompt: You are tasked with identifying the predominant moral narrative expressed by a text.

Prompt: Identify the predominant narrative expressed by the text, choosing between 'duty_to_care' ({dutycare}), 'duty_to_fight' ({dutyfight}), 'duty_to_educate' ({dutyeducate}), or 'other' if you cannot find an answer. DO NOT reply using a complete sentence, and ONLY give the answer in the following format: LABEL.

Text: {text}

Answer:
\end{promptbox}

\begin{promptbox}{Narratives - agency-oriented}
System prompt: You are tasked with identifying the predominant moral narrative expressed by a text.

Prompt: Identify the predominant narrative expressed by the text, choosing between 'right_to_good_health' ({rightogoodhealth}), 'right_to_inner_peace' ({rightoinnerpeace}), 'right_to_choose' ({rightochoose}), or 'other' if cannot find an answer. DO NOT reply using a complete sentence, and ONLY give the answer in the following format: LABEL.

Text: {text}

Answer:
\end{promptbox}

\section*{Appendix C}

The following list contains selected examples within the top-50 comments with the highest relative frequency of collective action markers overall:
\begin{itemize}
    \item ``Vegan don't fight, vegans unite."
    \item ``Do you support factory farming?"
    \item ``Do you even fight bro."
    \item ``Don't do over action ok?"
    \item ``So what do we do?"
    \item ``Okay, what do I do?"
    \item ``I support this calm protests."
    \item ``Facebook: you're supporting actual genocide by supporting facebook."
    \item ``Actually we do this everyday here."
    \item ``What do I do about that?"
    \item ``How do you join random meetings?"
    \item ``So, if this guy decides to actively join the cause and become vegan, will you actively join and support his cause?"
    \item ``Join us eaters."
    \item ``How can I do what you do?"
\end{itemize}

\section*{Appendix D}
In this section, we will detail existing assets used in terms of their licenses and the computational resources needed for the analysis.

\subsection{Licenses}
Code snippets provided by Google Developers for the use of YouTube API are licensed under the Apache 2.0 License.
Python packages used to process YouTube data, namely \texttt{Youtube Transcript}, \texttt{langdetect} and \texttt{Whisper} are licensed under an MIT license, while 
\texttt{Llama-2-70B-chat} model employed in the mapping validation is an open-source LLM released under a commercial use license\footnote{\url{https://ai.meta.com/llama/license/}}. \texttt{Mformer} models used in the extraction of moral scores are shared on HuggingFace and licensed under an MIT License, while \texttt{S-BERT} model for the extraction of sentence embeddings is licensed under the Apache License 2.0. The collective action dictionary by \citet{smith2018after} is publicly available.

\subsection{Resources}
We utilized a V100 30GB GPU for both extracting moral foundations scores through \texttt{Mformer} and generating sentence embeddings via \texttt{S-BERT}. All other resources were deployed locally on an Apple M1 Pro machine with 8 cores and 16GB of RAM.

\end{document}